\documentclass[journal]{IEEEtran}

\usepackage{cite}
\usepackage{amsmath,amssymb,amsfonts}
\usepackage{algorithm}
\usepackage{graphicx}
\usepackage{textcomp}
\usepackage[english]{babel}
\usepackage[utf8]{inputenc}
\usepackage[noend]{algpseudocode}
\usepackage{xcolor}
\usepackage[caption=false,font=footnotesize]{subfig}
\usepackage{array,colortbl}
\usepackage{bm}
\usepackage{multirow}
\usepackage{soul}          
\usepackage{url}           

\newcommand{\meddiamond}{\diamond}

\def\BibTeX{{\rm B\kern-.05em{\sc i\kern-.025em b}\kern-.08em
    T\kern-.1667em\lower.7ex\hbox{E}\kern-.125emX}}

\begin{document}

\title{Innovations in Cardless Artificial Intelligence Banking: A Comprehensive
Framework for Cyber Secure and Fraud Mitigation using Machine Learning Algorithms}

\author{\IEEEauthorblockN{Md Israfeel,~\IEEEmembership{Member, IEEE}}\\
\IEEEauthorblockA{Dept.\ Electrical and Computer Engineering,
University of Central Florida, Orlando, Florida, USA\\
israfeel.us@gmail.com}}

\maketitle

\begin{abstract}
The advent of cardless Artificial intelligence (AI) banking heralds a paradigm
shift in the financial landscape, offering users unprecedented security and
convenience. This paper outlines a comprehensive framework designed to elevate
cyber security, introduce auto-generated virtual cards, and mitigate fraud risks
within cardless AI banking systems. The framework envisions the future banking
architecture, employing AI-powered data cryptography to create secure virtual
cards for seamless transactions. By emphasizing the importance of secure
communication channels, it ensures the integrity of financial activities among
the bank system, cardholders, and third-party vendors. AI-based authorization
methodologies play a pivotal role in authenticating each transaction while
proactively identifying potential fraud, showcasing the framework's efficacy in
fortifying cardless AI banking security. The initial approach, featuring an
AI-driven, feature-based banking system, ensures the generate of virtual cards
with encrypted data, minimizing information exposure and reducing fraud risks.
Integrating a machine learning algorithm adds an extra layer of protection
against potential fraudulent activities. In conclusion, this proposed framework
establishes a holistic cybersecurity and fraud mitigation paradigm in cardless
AI banking systems. Its implementation empowers financial institutions to address
security concerns tied to traditional banking, paving the way for a future
banking landscape that is not only fraud-free but also secure and convenient for
users.
\end{abstract}


\section{Introduction}
\label{sec:introduction}
\IEEEPARstart{I}{n} the digital age, bank cards are integral to financial
transactions, yet their widespread use has made them a prime target for
cybercriminals [1, 31] and also The FBI has pinpointed synthetic identity theft
as the burgeoning star among financial crimes in the United States [2, 32].
Global e-commerce fraud is on the rise, hitting \$41 million in losses in 2022
and predicted to surpass \$48 billion in 2023 [3, 23, 24]. Europe, particularly
Germany and France, faces significant risk, while North America leads with over
42\% of global e-commerce fraud. Latin America experiences a 20\% revenue loss
to fraud, with 3.7\% of eCommerce orders being fraudulent. The adoption of
machine learning, AI, risk-scoring, and behavioral analysis is recommended for
retailers to establish effective fraud prevention measures in response to the
evolving fraud landscape. The paper underscores the urgent need for proactive
strategies as the cumulative losses to online payment fraud globally are
projected to exceed \$343 billion by 2027 [3, 25], and the urgency for enhanced
security measures is evident.

\begin{figure}[htbp]
    \centering
    \includegraphics[width=9cm]{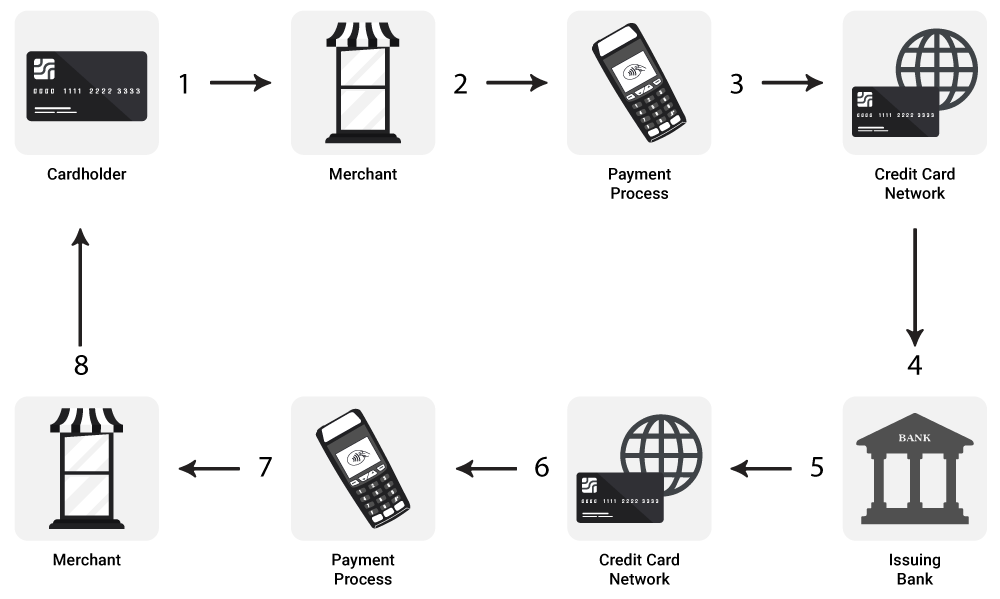}
    \caption{Current credit/debit card banking transaction overview.}
    \label{fig:fig1}
\end{figure}

\begin{figure*}[htbp]
    \centering
    \centerline{\includegraphics[width=1\textwidth]{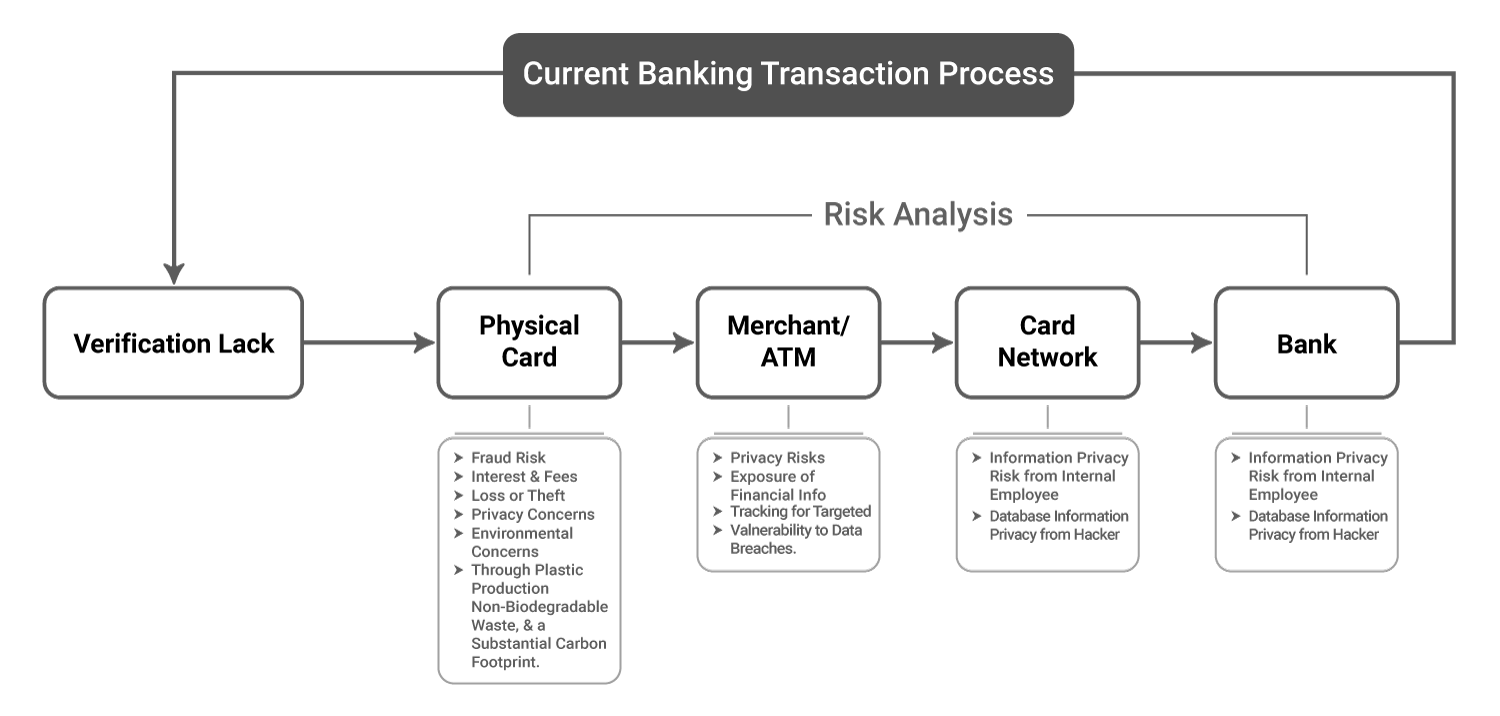}}
    \caption{Risk Analysis of Current Banking Card or credit/debit System}
    \label{fig:fig2}
\end{figure*}

This paper addresses the pressing research problem of developing comprehensive
security frameworks to combat credit/debit card fraud, protect sensitive
financial information, and mitigate risks in online transactions. Carrying
physical cards poses inherent disadvantages, including the risk of loss, theft,
or damage; potential unauthorized access leading to fraud; and the inconvenience
of physical bulk compared to digital alternatives. The surge in credit/debit
card fraud, exemplified by staggering statistics, underscores the critical need
for robust security measures. Prevalent tactics, including data breaches and
phishing scams, contribute to significant financial losses. Financial
institutions and technology providers are actively adopting strategies like
chip-based cards, multi-factor authentication, fraud detection systems, and
tokenization to bolster security. Consumer education on fraud prevention is also
essential. Figure 1 provides a snapshot of the existing banking card or
credit/debit processing transactions scenario.

In Figure 2, a visual representation exposes the inherent risk analysis
associated with the current process, emphasizing security vulnerabilities,
susceptibility to fraud, and concerns related to the bank, card network, and
data privacy. In this figure, we illustrate the intricate risk landscape of the
current banking card or credit/debit system, highlighting vulnerabilities at
various crucial stages. Physical cards as depicted in the diagram, face potential
threats such as theft, loss, and skimming devices, emphasizing the need for
enhanced security measures. Merchants: positioned as key intermediaries, are
portrayed in the context of data breaches, accentuating the inherent risks
associated with their role in the transaction process. The figure also delineates
the role of card networks emphasizing their susceptibility to cyberattacks that
may compromise data integrity during transactions. Simultaneously, the depiction
of banks underscores the challenge of identity theft, accentuating the importance
of robust verification processes and real-time monitoring to safeguard user
information. In navigating these risks, the figure visualizes strategic measures
such as advanced encryption, two-factor authentication, and continuous
monitoring, symbolizing the multi-layered approach adopted by the industry.
Furthermore, the representation emphasizes the critical role of customer
education as an essential component in fortifying the collective defense against
cyber threats and financial fraud. This visual representation in Figure 2 serves
as a comprehensive guide to understanding the nuanced risk dynamics within the
current banking card system, providing a valuable resource for stakeholders in
the financial industry to enhance their security protocols and strategies. The
FDIC's 2023 Risk Review, drawing insights from its unique perspective on
community banks, meticulously analyzes risks in the banking sector, including
credit, market, operational, crypto-asset, and climate-related financial risks
[26]. The report emphasizes the need for vigilance and strategic adaptation to
ensure the stability of the current banking card and credit/debit system in the
face of evolving challenges.

Moving forward, Figure 4 introduces a revolutionary bank card transaction system,
addressing current gaps and unveiling a novel approach to card processing.

In the rapidly evolving landscape of cardless AI banking, this research addresses
a critical problem: how to establish a comprehensive framework that navigates the
complexities of cyber security, creates secure virtual cards, and mitigates fraud
risks and merchant incidents for example The survey conducted in November and
December 2021 involved 1,060 merchants engaged in e-commerce fraud and payment
management. The participants were from four major geographic regions in North
America, constituting 41\% of the sample [4, 28] As traditional banking systems
face escalating threats in the digital era, the urgency to fortify defenses
against potential breaches and fraudulent activities becomes paramount. The
significance of this research extends to the broader context of credit card
security in the digital world. With global credit card fraud losses projected to
reach \$48 billion by 2023 [3, 25], the need for enhanced security measures is
evident. The adoption of new technologies, while providing convenience, has made
credit cards prime targets for cybercriminals, resulting in substantial financial
losses. Data breaches and phishing scams, common fraud tactics, contribute
significantly to this challenge. Financial institutions and technology providers
respond by implementing advanced security measures, including chip-based cards,
multi-factor authentication, fraud detection systems, tokenization, and consumer
education [5]. The research emphasizes the ongoing battle against credit card
fraud, aiming to contribute by exploring advanced security measures, raising
awareness that a physical banking card is not required for digital banking, and
fostering collaboration among industry stakeholders. In doing so, it seeks to
fortify the financial landscape against evolving cyber threats and ensure a
secure future for digital transactions.

As cyber threats evolve, the paper aims to contribute to the ongoing battle
against credit/debit card fraud by exploring and promoting advanced security
measures. By fostering collaboration among industry stakeholders, this research
seeks to ensure a secure and resilient financial landscape in the face of
evolving cyber threats.

This research delves into the transformative landscape of cardless AI banking,
addressing the critical need for a comprehensive framework to enhance
cybersecurity, introduce auto-generated virtual cards, and mitigate fraud risks.
Recognizing the urgency of defending against potential breaches and fraudulent
activities, the study sets forth specific objectives. Paperless Revolution: Ditch
the plastic, and embrace the digital future. Reduce environmental impact and
resource usage. 2. Fortress of Confidentiality: Build impregnable walls around
card data. Protect user information from everyone, including the user. 3.
Unbreakable Network: Forge a secure bridge between banks and networks. Ensure
seamless, encrypted communication, immune to data breaches. 4. Fraudsters
Beware: Design a system that sniffs out fraud before it happens. Real-time
monitoring and proactive measures keep your money safe. 5. User at the Helm:
Empower users with control over their finances. Every transaction requires
explicit authorization, with no exceptions. 6. Future-Proofed Security: Build a
system that adapts and evolves, staying ahead of ever-changing threats. Embrace
new technologies and maintain compliance with regulations. These objectives
collectively guide the research in equipping the financial landscape with
effective tools to combat evolving cyber threats and contribute to the evolution
of cardless AI banking into a secure, convenient, and resilient financial
ecosystem.

\section{Background}
In this segment, we delve into a myriad of investigations aimed at mitigating
global bank credit/debit card fraud. The endeavors span a spectrum of strategies,
encompassing fraud detection mechanisms, the implementation of multi-factor
authentication, the advent of visually fortified bank cards, and the exploration
of various survey-related studies. These investigations delve into the intricate
realms of financial crimes, the pervasive specter of global e-commerce fraud, and
the unsettling incidents of bank data breaches. The common thread weaving through
these diverse approaches is the meticulous monitoring of users' behaviors, an
intricate dance of scrutiny designed to eliminate any vestige of suspicion from
credit card transactions.

In a recent study conducted by Juniper Research and MasterCard, it was revealed
that global merchant losses to online payment fraud are projected to surpass
\$343 billion between 2023 and 2027. This staggering amount, exceeding 350\% of
Apple's reported net income for the fiscal year 2021, underscores the substantial
impact of online payment fraud on businesses worldwide, and in Figure 3, we
illustrate the total merchant losses from 2023 to 2028 [3, 25, 33]. Online
payment fraud encompasses losses incurred in various transactions, including
digital and physical goods sales, money transfers, and banking activities such as
airline ticket purchases. The methods employed by fraudsters encompass a range of
attacks, from phishing to business email compromise and socially engineered fraud.
The increase in online payment fraud losses is attributed, in part, to fraudster
innovation, particularly in areas like account takeover fraud, where perpetrators
hijack user accounts. Interestingly, these losses persist despite the widespread
implementation of identity verification measures. For a more in-depth exploration
of these emerging threats, segment analysis, and market forecasts in the realm of
online payment fraud, Juniper Research has published a comprehensive research
report titled ``Online Payment Fraud: Emerging Threats, Segment Analysis, and
Market Forecasts 2022--2027'' [33].

\begin{figure}[htp]
    \centering
    \includegraphics[width=9cm]{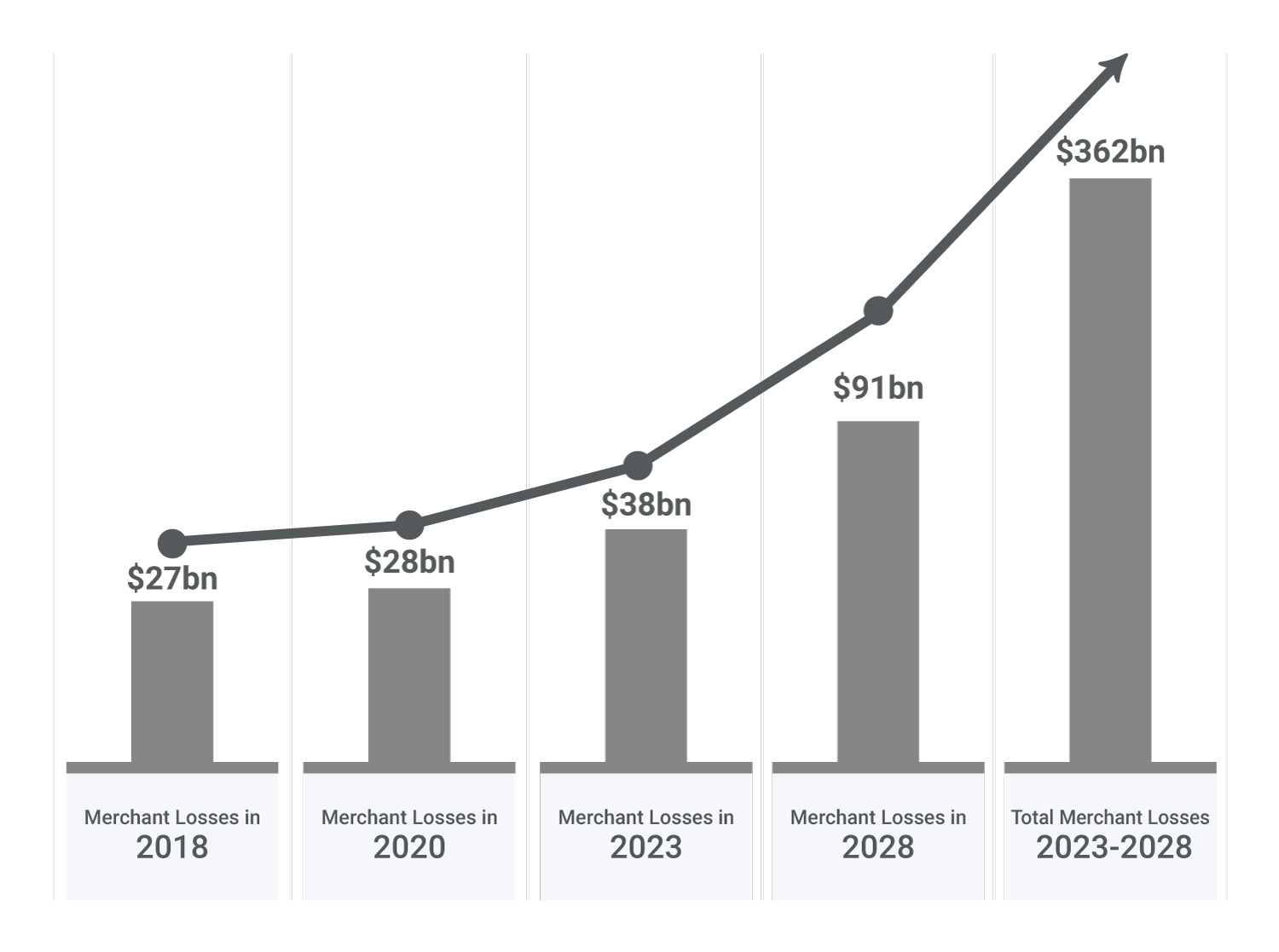}
    \caption{Analysis and Outlook: Online Payment Fraud Trends and Market
    Projections for 2023--2028.}
    \label{fig:fig3}
\end{figure}

Over the past two years, Canada has witnessed a surge in cybersecurity concerns,
underscored by frequent ransomware incidents disrupting essential services. The
National Cyber Threat Assessment (NCTA) 2023--24 identifies five dynamic
narratives shaping cyber threats until 2024, emphasizing the need for increased
vigilance and proactive measures [1]. Concurrently, the Money Laundering Risk
Assessment (NMLRA) acknowledges the transformative challenges posed by
cybercrime, ransomware attacks, and the opioid crisis within the U.S.\ financial
system. It accentuates the persistent vulnerabilities, including fraud, drug
trafficking, and compliance deficiencies, necessitating a comprehensive approach
to money laundering mitigation [2]. Furthermore, global e-commerce grapples with
an escalating fraud landscape, projected to exceed \$48 billion in losses in
2023. North America stands out with over 42\% of the global e-commerce fraud,
urging retailers to adopt advanced technologies like machine learning, AI, and
risk-scoring for robust fraud prevention strategies and enhanced financial system
resilience [3, 25].

In response to the escalating threat of fraud in credit card transactions, this
paper proposes an innovative online payment system to bolster security. The
system incorporates a secure communication channel, shared secrets, and
verification tokens, leveraging machine learning algorithms for generating
one-time credit card numbers on the user's device, a significant stride in
enhancing online credit card protection [6]. Addressing the surge in credit card
fraud, the study advocates for advanced machine learning techniques like the
Isolation Forest Algorithm (IFA) and Outlier Detection (OD) to detect and
prevent Master Card fraud transactions [7], utilizing a dataset of 284,807 card
transactions. The paper also highlights the substantial increase in payment card
fraud due to the growing use of payment cards for online purchases. A study
conducted in Slovakia through electronic sample surveys sheds light on opinions
and consumer approaches to security and payment card fraud, providing insights
for financial institutions to enhance fraud prevention measures [8].

The surge in smartphone and tablet usage has significantly impacted e-government
services. This paper proposes an advanced authentication scheme integrating
multi-factor authentication, including a One-Time Password (OTP) and biometric
authentication. This approach enhances security, providing a robust and
user-friendly solution for accessing e-government services [9]. The proposed
system introduces a unique approach to enhance credit card security by allowing
the dynamic modification of the Card Verification Value (CVV). Developed by
J.M.~Ashfield [10], this method utilizes at least a portion of a one-time
password to generate a dynamic CVV for prepaid, debit, or credit cards. Unlike
traditional CVVs, these dynamic values have limited use, specifically designed
for online transactions. Users can gain authorization through a dynamic CVV by
receiving a transaction authorization request that includes this code. The system
then compares the dynamic CVV to a one-time password generated for the specific
credit/debit card, enhancing security for each online transaction. Successful
matches result in transaction authorization sent to the merchant, ensuring a
robust and dynamic security framework.

Previous work has predominantly focused on mitigating the challenges posed by
traditional authentication methods, particularly passwords. Researchers have
explored alternatives, including multi-factor authentication and biometric
solutions, to bolster security. Efforts have also addressed the increasing use of
smartphones and tablets in accessing e-government services. Building upon past
work, this research addresses persisting challenges in authentication methods,
specifically weaknesses associated with traditional passwords. The proposed
advanced authentication scheme integrates multi-factor authentication and
biometrics, contributing to a more robust and user-friendly authentication
process for secure access to e-government services. Following this, we will delve
into a comprehensive discussion of the proposed system.

\subsection*{Related Work}
In response to the escalating threat of cyberattacks and credit card fraud in the
banking sector, this study critically assesses existing approaches, highlighting
the shortcomings of conventional methods and emphasizing the growing role of
machine learning. The review, encompassing 181 research articles from 2019 to
2021, meticulously explores machine learning and deep learning techniques for
credit card cyber fraud detection. Providing valuable insights, the review aids
researchers and the banking industry in selecting effective strategies, addressing
current challenges, and outlining future research directions for robust cyber
fraud detection innovation projects [27].

The escalating threat of cyberattacks has positioned cybersecurity as a critical
priority in the banking sector, with credit card cyber fraud emerging as a major
global security concern. Traditional approaches like anomaly detection and
rule-based techniques, while prevalent, suffer from inefficiencies. In response,
machine learning (ML) has become instrumental in revolutionizing cyber fraud
detection. This study conducts an exhaustive review, analyzing 181 research
articles from 2019 to 2021, focusing on ML and deep learning (DL) applications
in credit card cyber fraud detection. The goal is to furnish researchers and the
banking industry with valuable insights for innovative cyber fraud detection
projects [28]. The study's conclusions underscore the necessity for further
exploration of semi-supervised and unsupervised learning techniques, with a
specific recommendation for deep learning approaches in credit card cyber fraud
detection. Advocating for the integration of ML and DL algorithms to bolster
detection outcomes, the study calls attention to the challenges of imbalanced
datasets, proposing both oversampling and undersampling techniques. Researchers
are encouraged to transparently report dataset sources and performance metrics,
fostering consistency and clarity. Additionally, the banking industry is urged to
contribute diverse datasets of fraudulent activities for collaborative research
initiatives. This comprehensive review aims to catalyze innovation in cyber fraud
detection systems, benefiting both researchers and the banking sector.

\section{Proposed System}
The proposed system introduces an innovative phase, as depicted in Figure 4,
adding an extra layer of security before engaging in standard transactions using
conventional equipment. Before initiating any banking transaction via the card,
the account user is required to obtain access from the bank, logging in through a
smartphone or personal computer using their unique login credentials.

\begin{figure}[htp]
    \centering
    \includegraphics[width=8.5cm]{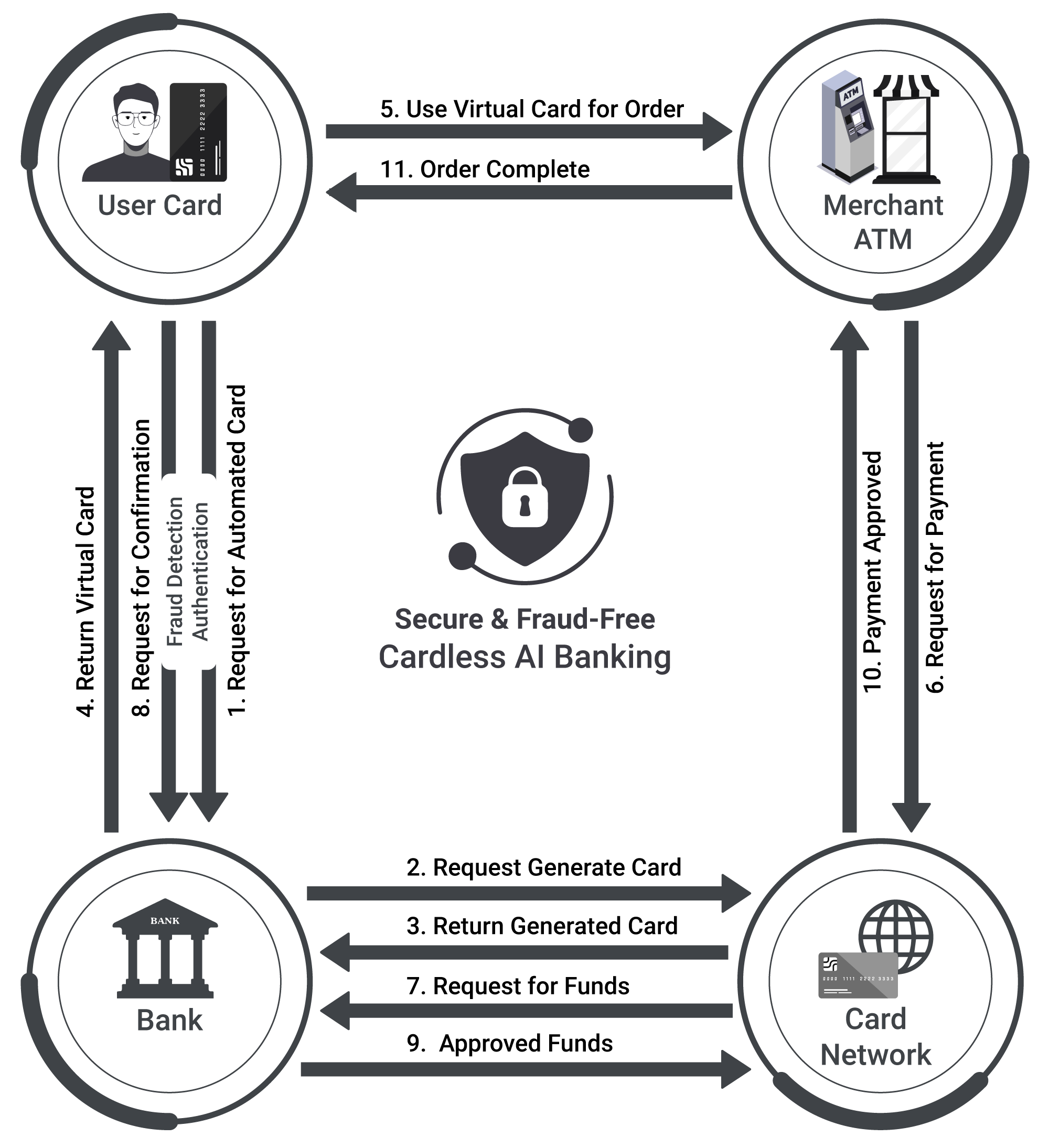}
    \caption{The proposed banking card transaction system}
    \label{fig:fig4}
\end{figure}

In the initial step of this novel phase (Phase 1), the user initiates a request
for an automatic card generate from both the bank and card network. This step
involves various policies that govern the card's functionality, such as
determining whether it's for one-time or multiple uses, and specifying the
generated card's monetary limits. Details of these policies are thoroughly
outlined during this first step. Upon the user's request for card generate, the
bank activates the card generator, prompting a request to the card network
(Phase 2) for virtual card creation. The deployment section provides a detailed
description of the policies guiding this card generate process. The card,
generated by the network, is then transmitted back to the bank (Phase 3) and
sent to the user's encrypted bank card (Phase 4), which subsequently forwards it
to the account user (Phase 5). The user gains the ability to use this card for
various purposes, such as online shopping or ATM withdrawals. Importantly, the
merchant only becomes aware of the card network code during any transaction,
ensuring the confidentiality of the actual card details.

When the user presents the card to the merchant (Phase 6), the merchant, having
access only to the card network number, forwards the payment request to the
network. The card network decodes the card details, validates its authenticity,
and prompts a fund request to the bank (Phase 7). Leveraging fraud detection and
authentication technology, including AI, the bank verifies the user and seeks
confirmation for the transaction specifics. Upon receiving positive confirmation
from the user (Phase 8), the network proceeds to accept the user's order,
ultimately confirming payment receipt to the merchant (Phases 9, 10, 11). This
dynamic system not only enhances transaction security but also distinguishes
between moderate and risky transactions based on user spending behavior. It
employs advanced user-server communication to preemptively address potential
fraudulent activities, establishing a robust and secure transaction environment.

\section{Methodology}
This study adopts a robust methodology aimed at bolstering the security and
efficacy of the banking card system. It entails a multifaceted approach,
encompassing several critical components such as bank card generate, homomorphic
encryption implementation for information secure, machine learning-based fraud
detection, and the optimization of payment verification procedures. Each facet is
crucial in enhancing the overall security and efficiency of banking card
transactions. By meticulously addressing these key aspects, this research
endeavors to contribute to the advancement of secure and reliable banking card
systems.

\subsection{Process of the Bank Card Generate}
Generating a credit/debit card number involves multiple components. Payment card
numbers typically range from 8 to 19 digits in length. The initial six or eight
digits constitute the issuer identification number (IIN), also known as the bank
identification number (BIN). Following the IIN, the subsequent digits, excluding
the last one, denote the individual account identification number. The final
digit serves as the Luhn check digit, ensuring the accuracy of the card number.
Both the IIN and PAN (Primary Account Number) adhere to a standardized numbering
scheme established by ISO/IEC 7812. The structure of these numbers includes: (a)
A six or eight-digit IIN, with the first digit representing the major industry
identifier (MII). (b) A variable-length individual account identifier, spanning
up to 12 digits. (c) A single check digit computed using the Luhn algorithm,
validating the integrity of the card number. Including the Issuer Identification
Number (IIN), and individual account identifier (the system generates a
distinctive 9-digit number as a cardholder identifier. Importantly, the system
avoids reissuing the same number for any subsequent online transactions until it
has been utilized and discarded.), and a check digit. The Luhn algorithm is
commonly used to validate credit card numbers [34]. Below is a simplified
explanation of generating a credit card number:

Issuer Identification Number (IIN): Define a specific IIN based on the credit
card issuer (e.g., Mastercard, Visa). Let the IIN be denoted as \textit{I}, and
it typically consists of six digits. Individual Account Identifier: Generate a
unique 9-digit individual account identifier ($I_\mathrm{ID}$) using a specific
algorithm or approach. Check Digit: Concatenate the IIN and $I_\mathrm{ID}$ to
form a 15-digit number. Use the Luhn algorithm to calculate the check digit
(\textit{C}):
\begin{equation}
C = (10 - (S \bmod 10))
\end{equation}
where \textit{S} is the sum of digits obtained by doubling every second digit
from the right in the 15-digit number. Bank Card Number: Combine the IIN,
$I_\mathrm{ID}$, and the calculated check digit to form a valid 16-digit credit
card number. Mathematically, the bank card number (\textit{BCN}) can be
represented as (where $\|$ denotes concatenation):
\begin{equation}
BCN = I \;\|\; I_\mathrm{ID} \;\|\; C
\end{equation}
In the process of generating a card number, to guarantee exclusivity, the server
scrutinizes the one-time number for duplication within the database before
activation, ensuring that the generated number hasn't been previously stored in
the server's database. This meticulous checking process prevents any inadvertent
repetition of generated numbers, reinforcing the security of the card system.

\subsection{Encryption Bank Information}
The digital revolution in finance has intensified the need for robust security in
online transactions, especially in the banking card transaction system dealing
with sensitive data like credit card numbers. Encryption techniques have become
essential for safeguarding financial data and ensuring its confidentiality,
integrity, and authenticity. Acting as a protective barrier, encryption renders
information indecipherable to unauthorized entities during storage and
transmission, which is crucial in combating persistent threats like data breaches.

\subsubsection{Homomorphic Encryption}
Homomorphic Encryption, a significant advancement in banking security, allows
computations directly on encrypted data, offering secure processing without
compromising privacy [29]. This innovative approach addresses critical challenges
in the digital banking arena, providing an additional layer of security during
computations on encrypted data. Homomorphic encryption represents a significant
evolution in cryptographic methods, distinguishing itself by enabling usability
with encrypted data. This unique feature allows for calculations and analyses to
be conducted directly on encrypted data [19, 18]. The term ``homomorphic''
originates from the Greek words ``homos'' and ``morphe,'' collectively meaning
``same shape'' [17]. In algebraic terms, a homomorphism refers to a
structure-preserving mapping between two algebraic structures [16]. In the
context of cryptography, this implies that the essential properties of an
unencrypted dataset are retained and transferred to the encrypted dataset during
the encryption process [18].

\begin{figure}[htp]
    \centering
    \includegraphics[width=9cm]{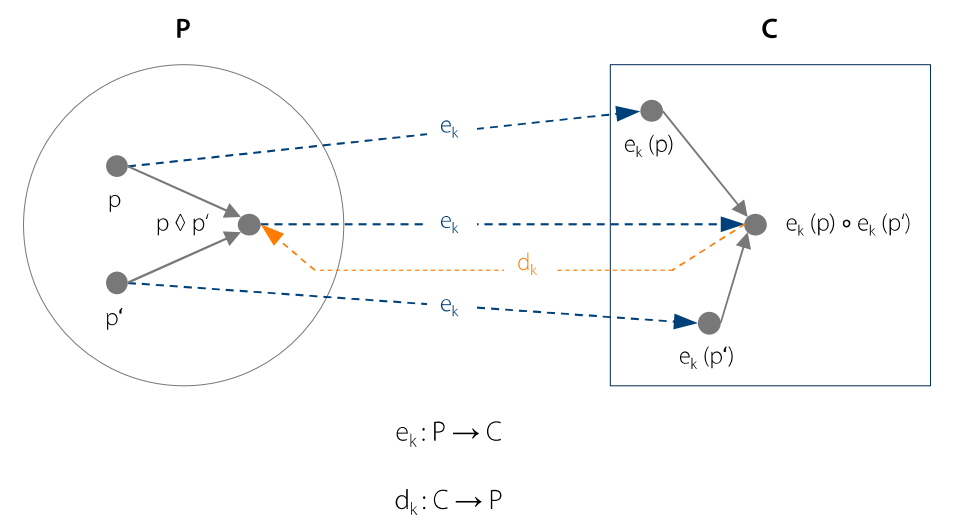}
    \caption{Homomorphism of an encryption function [35].}
    \label{fig:fig5}
\end{figure}

Figure 5 illustrates the homomorphism of encryption within the introduced
quintuple of cryptosystems. Drawing parallels to mathematical homomorphism, the
encryption (Equation~(3)) and decryption (Equation~(4)) are defined analogously.
\begin{equation}
e_k(p \meddiamond p') = e_k(p) \circ e_k(p') = c \circ c',
\end{equation}
\begin{equation}
d_{k}(c \circ c') = d_k(c) \meddiamond d_{k}(c') = p \meddiamond p',
\end{equation}
where $\meddiamond \in \{\bigotimes, \bigoplus\}$, $\forall\, p, p' \in P$;
$e_k$ is an encryption function, $d_k$ is a decryption function, $p, p'$ is
plaintext, $c, c'$ is ciphertext, $\oplus$ denotes addition, and $\bigotimes$
denotes multiplication.

Exploring the intricacies of encryption in banking card transactions, this
section highlights the indispensability of protective measures and delves into
the distinctive features and applications of Homomorphic Encryption. The
deployment of advanced encryption techniques plays a pivotal role in securing
bank information details and mitigating risks associated with unauthorized access
and data breaches. This practice is fundamental in ensuring the integrity of
electronic transactions or stored data and fostering trust among users, financial
institutions, and merchants in the digital banking ecosystem.

\subsection{Fault Transaction Detection Using ML}
Fault detection is critical for maintaining the security and reliability of card
bank systems in the digital finance landscape. This process involves identifying
irregularities, technical glitches, or potential fraudulent activities that could
disrupt the normal functioning of the system. The timely detection of faults is
essential to prevent unauthorized transactions, protect sensitive user
information, and minimize financial losses. Advanced fault detection mechanisms
often utilize machine learning and artificial intelligence to analyze transaction
patterns and user behavior, enabling the system to learn from historical data and
identify anomalies. In summary, robust fault detection is integral to upholding
the integrity of card bank systems, ensuring user trust, and facilitating the
smooth operation of digital banking ecosystems.

Logistic Regression (LR) is a valuable machine-learning algorithm for fraud
detection based on account history. Leveraging historical data as features, LR
excels in binary classification tasks, distinguishing between fraudulent and
non-fraudulent transactions. Trained on labeled data, the LR model learns
patterns and establishes a decision boundary to make real-time predictions. Its
interpretability and adaptability make it suitable for evolving fraud patterns.
The importance of features is revealed through coefficients, aiding in
understanding and explaining predictions. The logistic regression model predicts
the probability ($p$) of a transaction being fraudulent using the logistic
function:
\begin{equation} \label{eq:lr}
p(x) = \frac{1}{1 + e^{-(\beta_0 + \beta_1 x_1 + \beta_2 x_2 + \dots + \beta_n x_n)}}
\end{equation}
Here, $p(x)$ is the probability of fraud; $\beta_0, \beta_1, \ldots, \beta_n$
are coefficients learned during training; and $x_1, x_2, \ldots, x_n$ represent
the features of the account history data. The decision boundary is determined by
a threshold (commonly 0.5). If $p(x)$ exceeds this threshold, the transaction is
predicted as fraudulent; otherwise, it is predicted as non-fraudulent.

\subsection{Transaction Verification}
Transaction verification from the user is a crucial step in ensuring the security
and integrity of transactions, particularly in online banking and financial
systems. This process involves confirming the identity of the user to prevent
unauthorized access and fraudulent activities. Various verification methods are
employed, including knowledge-based factors like passwords or PINs,
possession-based factors such as mobile authentication or smart cards, and
biometric factors like fingerprints or facial recognition. Implementing
multi-factor authentication enhances the robustness of user verification, adding
layers of security to sensitive transactions and account access. The mathematical
representation of a simple knowledge-based verification process can be expressed
as follows:
\begin{equation}
\mathrm{UserVerify} = \mathrm{Fun}(\mathrm{UserInput},\, \mathrm{StoredCredentials})
\end{equation}
Here, the function compares the user-inputted information (passwords or PINs,
possession-based factors such as mobile authentication or smart cards, and
biometric factors like fingerprints or facial recognition) with the stored
credentials. If there is a match, the user is verified successfully, granting
access to the intended transaction or system.

\section*{Deployment Overview}
In the ever-evolving landscape of digital finance, securing bank credit/debit
card transactions has become a paramount concern. Our innovative approach
introduces an additional phase to the standard transaction process, bolstering
security measures. Before users engage in routine transactions, they must have
customized client bank software installed or log in through our dedicated
website.

\begin{figure}[htp]
    \centering
    \includegraphics[width=9cm]{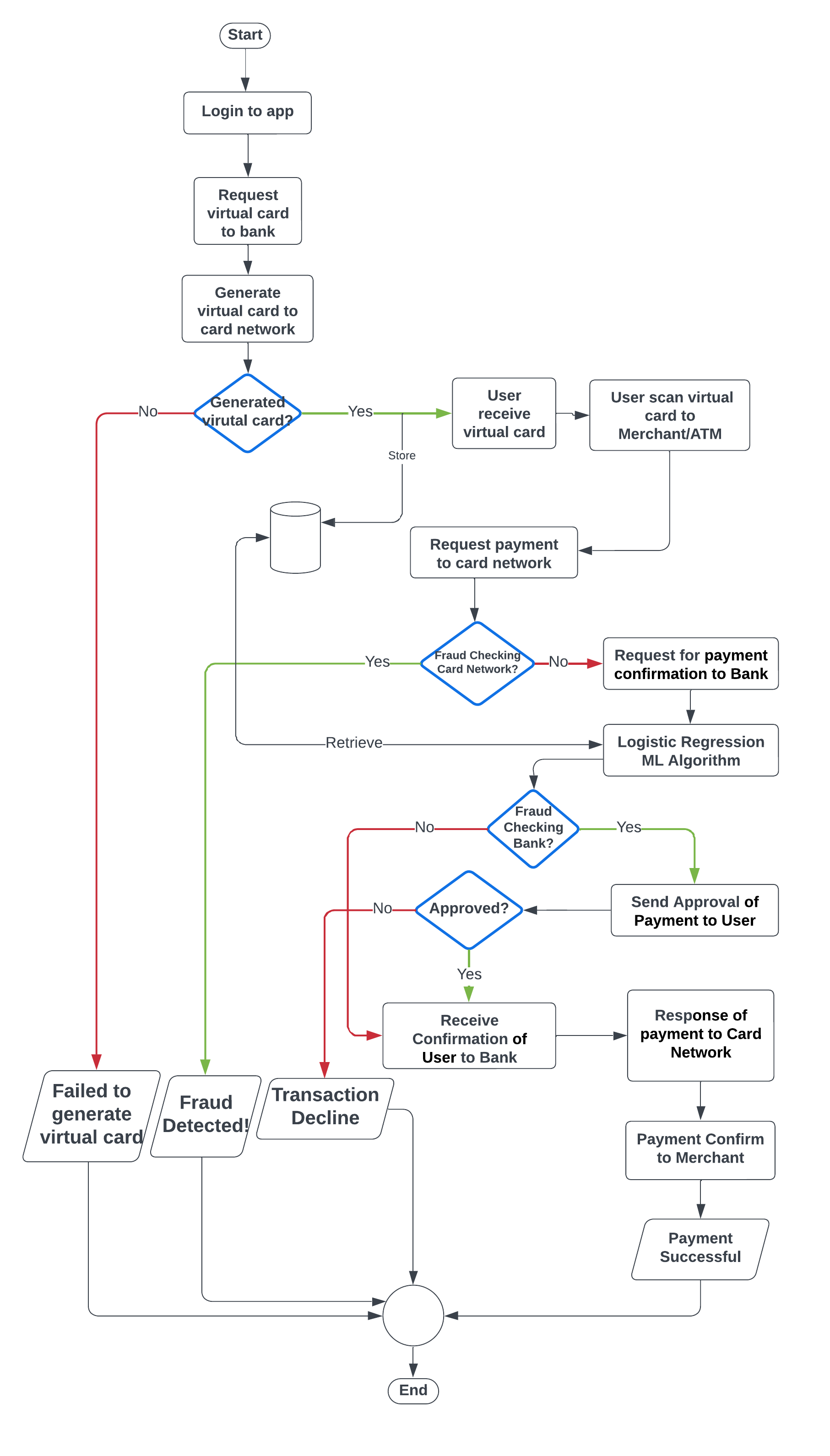}
    \caption{Flowchart of Bank Card Transactions Process with Machine Learning
    (ML) Integration.}
    \label{fig:fig6}
\end{figure}

This deployment overview delves into the intricate steps and key elements
involved in this enhanced bank card transaction system. 1.~User Interaction and
Identity Verification: Users play a pivotal role in this deployment. They
interact with the system using smartphones or personal computers. To access the
generated card number, users provide transaction details such as the amount and
validity period. Identity verification is a critical step, requiring users to
enter their username and password for any transaction. 2.~Bank Operations: The
bank serves as the hub for most operations. It accepts virtual card generate
requests from users, acting as the central point for transaction processing.
Server-side operations include identity verification, ensuring the security and
legitimacy of the user. 3.~Card Network Functionality: The Card Network comes
into play upon receiving a request from the bank for virtual card generate. It is
responsible for providing virtual cards to the bank securely, forming a crucial
link in the transaction chain. 4.~Virtual Card Generate and Encryption: The Card
Network generates virtual cards, and the bank encrypts these before being sent to
users. The encrypted virtual cards become the user's key for secure transactions.
5.~User Interaction with the Virtual Card: Users utilize the encrypted virtual
cards for transactions, whether it be making payments at a merchant's point of
sale or withdrawing cash from an ATM. 6.~Transaction Processing at
Merchant's/ATM: Once the user initiates a transaction, the merchant or ATM
processes the request. The card network steps in to confirm payment requests,
adding an extra layer of validation. 7.~Machine Learning-Driven Security Checks:
The card network employs machine learning algorithms to check the validity of the
card and detect any signs of fraud. This proactive measure enhances security and
minimizes the risk of unauthorized transactions. 8.~Verification of Fund
Availability: A critical checkpoint in the process involves checking fund
availability. The card network ensures that the user has sufficient funds for the
proposed transaction before proceeding. 9.~User Authorization and Transaction
Approval: Machine learning algorithms play a pivotal role in authorizing users.
The system conducts comprehensive checks to ensure that the user is legitimate
and approves transactions accordingly. 10.~Fund Approval: After the rigorous
checks, the card network obtains approval for the payment from the bank,
confirming that the funds are available and the transaction is legitimate.
11.~Payment Successful: Upon receiving approval, the card network confirms the
successful payment to the merchant or ATM. This finalizes the transaction,
providing a seamless and secure experience for the user.

In delineating our novel banking system, we have elucidated the process through a
comprehensive flowchart seamlessly integrated with Machine Learning (ML)
techniques, as illustrated in Figure~6. This visual representation encapsulates
the intricate steps involved in our innovative approach. This deployment overview
encapsulates the orchestrated series of steps and interactions among users,
banks, and card networks, each contributing to the enhanced security of credit
card transactions. The integration of machine learning adds a layer of
intelligence, making the system adept at identifying and preventing potential
fraudulent activities. As the digital financial landscape continues to advance,
this deployment represents a forward-looking solution that prioritizes both
security and user experience.

\begin{algorithm}
\caption{Secure virtual card generate and transaction for bank system}
\label{alg:cap}
\begin{algorithmic}[1]
\State \textbf{Input: Card payment process}
\State \hspace{8mm} User credentials $X = (x_1, x_2)$
\State \hspace{8mm} Define card policy $Y = (y_1, y_2, y_3, \dots, y_n)$
\State \hspace{8mm} Card generate $G = (X, Y)$
\If{$G \neq \emptyset$}
  \State Card stored in the card network $N \gets G$
  \State User receives virtual card $X \gets V$
  \State Scan virtual card to merchant/ATM $M$
  \State \textsc{PaymentRequest}$(V, M)$
  \If{\st{$F = \emptyset$}} \Comment{$F = \text{Fraud}$}
    \State \textsc{RequestPaymentConf}$(V, B)$ \Comment{$B = \text{Bank}$}
    \If{$F < 0.5$} \Comment{Fraud detected via LR and ML}
      \State \textsc{PaymentApproval} $X = (V)$
      \If{\st{\textsc{UserApproved}() $\in \mathcal{T}$}}
        \State Response to network $N \gets B \gets X$
        \State Notify user $X \gets M \gets N$
        \State \textbf{return} Payment completed successfully!
      \Else
        \State \textbf{return} User approval failed!
      \EndIf
    \Else
      \State \textbf{return} Fraudulent transaction!
    \EndIf
  \Else
    \State \textbf{return} Fraud detection failed!
  \EndIf
\Else
  \State \textbf{return} Virtual card generate failed!
\EndIf
\end{algorithmic}
\end{algorithm}

Furthermore, we have introduced Algorithm~1: Secure virtual card generated and
transaction for the bank card system. This algorithm encapsulates a secure
methodology for the generate of virtual cards within our system. Through a
meticulous sequence of steps, it ensures the robust generate of secure virtual
cards, embodying the core principles of our advanced banking framework. Together,
the flowchart and algorithm provide a clear and concise overview of our approach,
harmonizing visual clarity with procedural details. These tools serve as
fundamental components in understanding the intricacies of our new banking
paradigm, underscored by the integration of machine learning for enhanced
security and efficiency.

The proposed system utilizes the existing infrastructure and network security
systems such as SSL. Users must create accounts, and account-encrypted
information is stored in the system's database for future login and verification
purposes. Additionally, Figure~6 depicts the flow charts for virtual bank card
number uniqueness verification and the integration of machine learning (ML) for
fraud detection based on the user's transaction history. Credit/Debit card number
generate: According to PCI DSS, the 16-digit credit card number is structured
with the first six digits identifying the card issuer and the following 7--15
digits serving as the cardholder identifier. The system generates a unique 9-digit
cardholder identifier and ensures non-reissuance for any online transaction until
the number is used and discarded. The last digit is a check number generated
using the Luhn algorithm for accuracy. The proposed system enhances security with
three dimensions: a virtual credit card number, a secure communication medium,
and integration with an ML fraud detection algorithm. The generated credit card
numbers are stored in the database for user card policies, and after this period,
they are no longer valid for future transactions. This multi-layered approach
ensures high security for every online credit/debit card transaction.

\begin{figure}[htp]
    \centering
    \includegraphics[width=9cm]{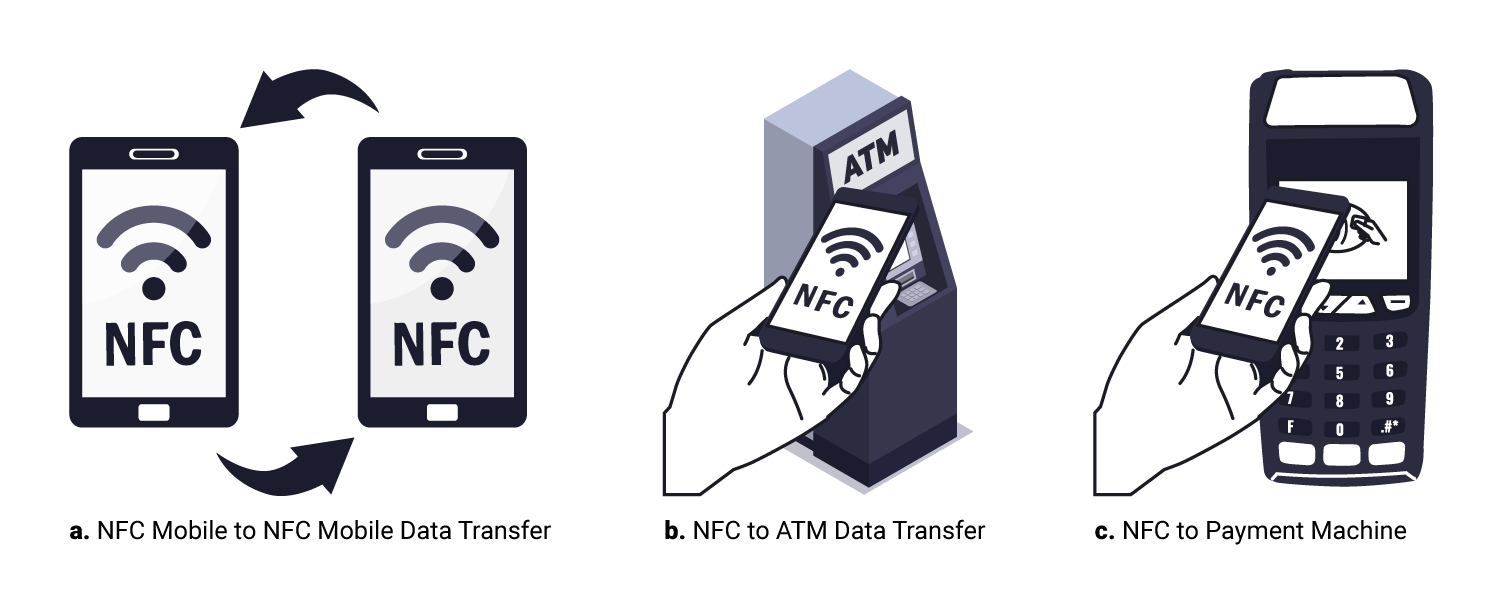}
    \caption{NFC to ATM, NFC to POS, or NFC to NFC}
    \label{fig:fig7}
\end{figure}

In the evolving digital payment landscape, the Near Field Communication (NFC)
technology paradigm has paved the way for a seamless and secure transaction
experience. This transformative approach replaces the conventional use of
physical cards with the convenience of NFC-enabled mobile devices, offering users
a multifaceted range of transaction possibilities. Figure~7 illustrates the
versatility of NFC payments, showcasing three key scenarios: Figure~7(a) NFC to
NFC: The NFC technology facilitates peer-to-peer fund transfers between users'
mobile devices. After completing a payment transaction, users can seamlessly
transition into an NFC-enabled fund transfer setup. This enables users to
initiate secure transfers to other NFC-enabled devices within close proximity.
The process involves authentication, transfer confirmation, recipient
authorization, and the successful completion of the NFC fund transfer.
Figure~7(b) NFC to ATM: Users can leverage their NFC-enabled mobile devices to
interact with ATMs securely. By initiating transactions through their
smartphones, users can seamlessly withdraw cash, and check balances, and perform
various ATM-related operations without the need for a physical card. The secure
communication between the mobile device and the ATM ensures a reliable and
user-friendly experience. Figure~7(c) NFC to Merchant: NFC-enabled mobile
payments extend to merchant transactions, enabling users to purchase at
point-of-sale terminals. Users can effortlessly tap their mobile devices on
NFC-equipped merchant terminals, initiating secure and swift transactions. This
contactless method enhances the overall payment experience for both consumers and
merchants, fostering efficiency and reducing the reliance on traditional payment
cards. This functionality not only enhances convenience but also exemplifies the
versatility of NFC technology in fostering a mobile-centric, cardless financial
ecosystem. The integration of NFC technology in these scenarios not only
simplifies and expedites transactions but also contributes to the ongoing
evolution of digital finance. The secure and contactless nature of NFC
transactions aligns with the growing demand for efficient and user-centric
payment solutions, ultimately shaping the future of the financial landscape.

\begin{figure}[htp]
    \centering
    \includegraphics[width=8cm]{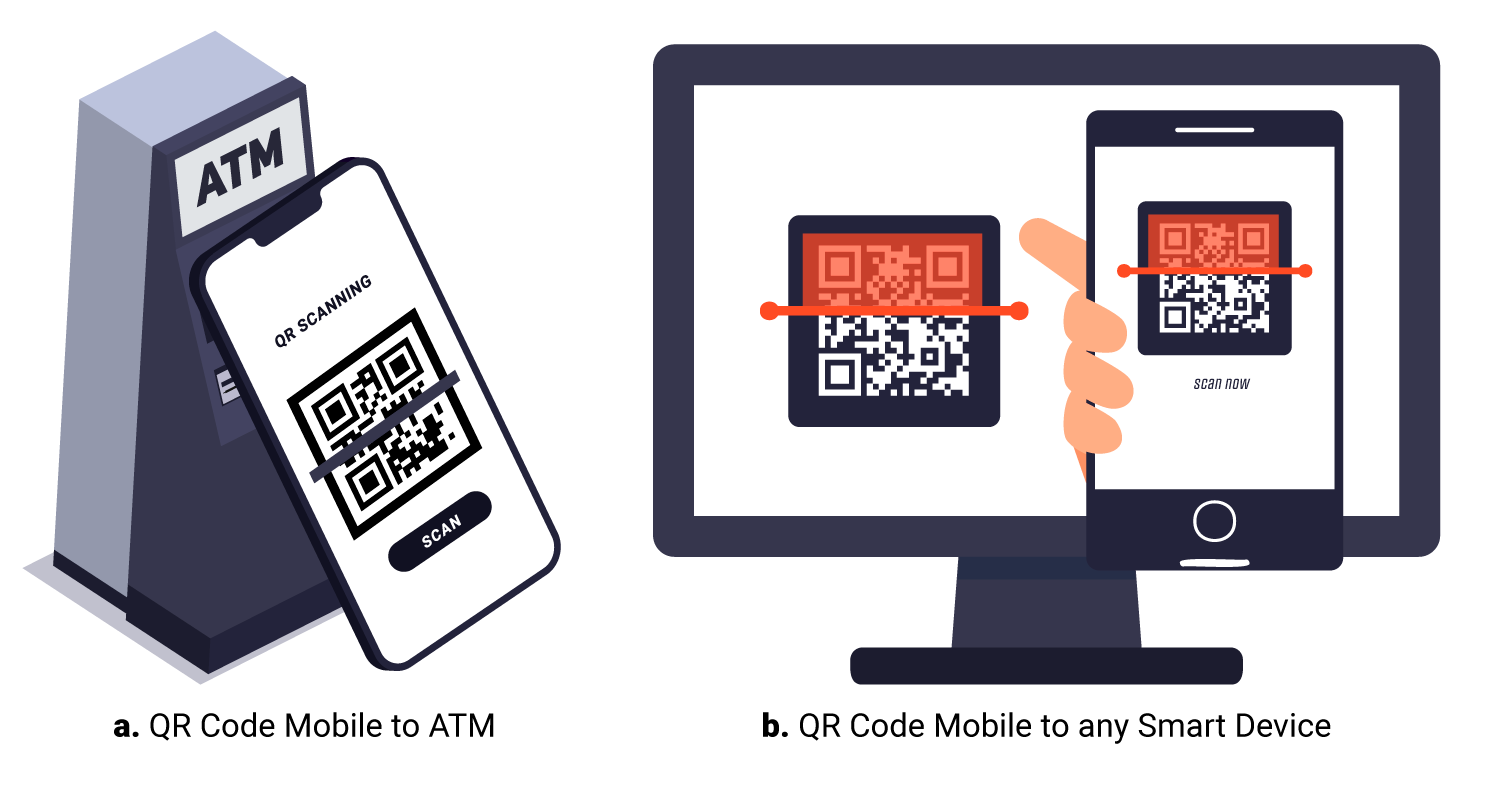}
    \caption{QR code for any smart device}
    \label{fig:fig8}
\end{figure}

Figure~8 illustrates the versatility of virtual credit/debit card payments. In a
dynamic shift towards modernized and convenient payment methods, QR codes have
become a versatile tool for facilitating transactions on any smart device. This
innovative approach enables users to embrace a seamless and secure payment
experience, transcending the limitations of traditional physical cash. Users can
leverage QR codes generated on their smart devices to represent virtual secure
cards. Whether on a smartphone, tablet, or any other smart device, these QR codes
serve as dynamic and secure conduits for initiating transactions. The user-friendly
nature of QR codes ensures compatibility across a wide range of smart devices,
providing a universal and accessible solution for digital transactions. This
inclusive use of QR codes on any smart device enhances convenience and aligns
with the contemporary emphasis on contactless and efficient payment methods. It
empowers users to engage in transactions with ease, offering a flexible and
modernized alternative to traditional payment mechanisms. The versatility of QR
codes, coupled with the ubiquity of smart devices, contributes to a transformative
shift in the landscape of digital finance.

\begin{figure}[htp]
    \centering
    \includegraphics[width=5cm]{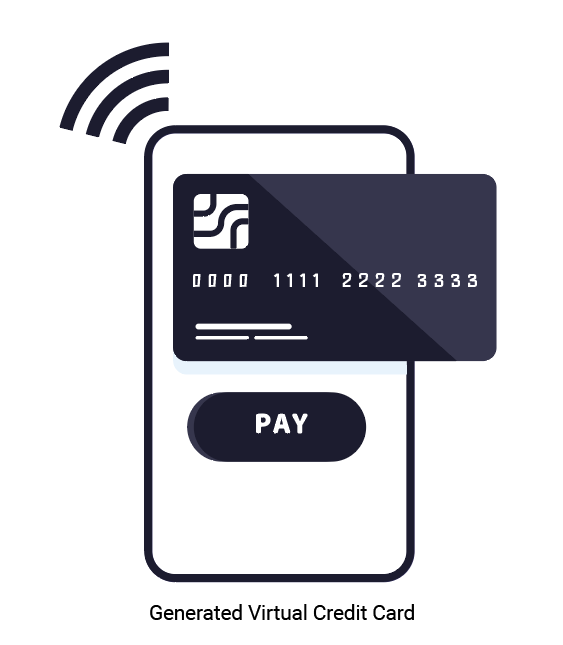}
    \caption{Virtual one-time credit/debit card}
    \label{fig:fig9}
\end{figure}

Figure~9 illustrates the virtual secure card, which presents a revolutionary
approach to payment methods, offering users the flexibility to wield it much like
a conventional physical card. This innovative solution not only preserves the
familiarity of traditional transactions but also introduces newfound adaptability
and convenience. Users have the autonomy to utilize the virtual secure card in a
manner akin to a manual card, enabling them to make purchases, withdraw funds,
and conduct various transactions seamlessly. The transition from physical to
virtual is seamless, allowing users to retain the tactile experience of handling a
card while embracing the enhanced security and features embedded in the digital
realm. Furthermore, the virtual secure card extends its utility by providing users
with the option to share it with their economic counterparts. This collaborative
feature fosters a networked approach to financial interactions, allowing users to
extend the benefits of secure and convenient transactions to their economic peers.
The virtual secure card, with its versatile functionality, emerges as a dynamic
tool that not only modernizes individual transactions but also facilitates a
connected and efficient financial ecosystem.

Opting for a virtual card over its physical counterpart not only enhances
security but also expands the scope of its applications. The transition to virtual
cards introduces heightened security measures and versatility, offering users a
more secure and flexible means of conducting transactions. This shift reflects the
evolving landscape of financial technology, emphasizing the adaptability and
enhanced features that virtual cards bring to the forefront of modern payment
methods.

\section*{Results}
The realm of e-commerce fraud is on the rise, evidenced by a substantial increase
in losses from \$41 million in 2022 to a projected surge exceeding \$48 billion
in 2023, as depicted in Figure~10. These figures underscore the escalating
challenges faced by the credit card banking sector in combating projected losses
attributed to fraud [22]. Notably, North America leads in fraudulent transaction
value, accounting for 42\% globally, while Europe, specifically Germany and
France, faces substantial e-commerce fraud risks. Latin America experiences a
20\% revenue loss to fraud, and Asia Pacific witnesses a severe challenge with a
cost of \$4 per fraudulent transaction.

\begin{figure}[htp]
    \centering
    \includegraphics[width=8.5cm]{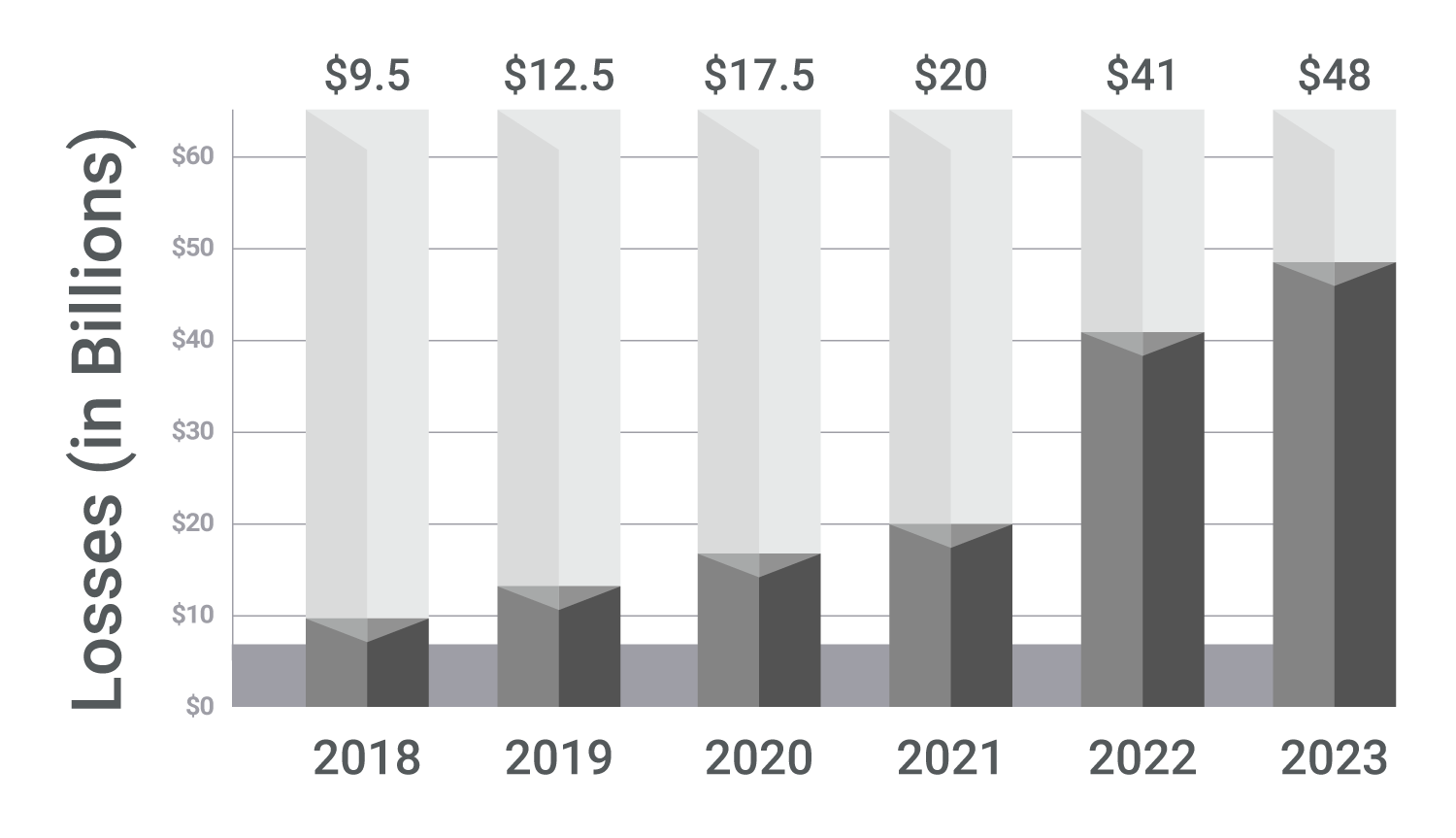}
    \caption{Credit Card Banking Projected Losses From Fraud}
    \label{fig:fig10}
\end{figure}

To combat this, merchants are advised to integrate advanced technologies like
machine learning, AI, risk-scoring, and behavioral analysis into their systems
for comprehensive fraud prevention [1]. The statistics convey a disconcerting
narrative in the realm of identity theft and credit card fraud. Despite a
marginal decrease in reports, these types of fraud have persisted since 2020,
surpassing pre-pandemic levels into the first three quarters of 2023. Following a
doubling trend between 2019 and 2020, identity theft reports continued to rise in
2021, impacting nearly 1.4 million individuals. The Federal Trade Commission
(FTC) collected approximately 1.1 million reports in 2022, with 805,000 reports
filed through the first three quarters of 2023. In 2022, reports of identity
theft reached 1.108 million, with 805,000 instances reported from January to
September 2023 [21]. We present Table~1, showcasing credit card fraud reports
categorized by year.

\begin{table}[h]
    \centering
    \caption{\textbf{Credit card fraud reports by year}}
    \setlength{\tabcolsep}{6pt}
    \renewcommand{\arraystretch}{1.5}
    \begin{tabular}{|c|c|}
        \hline
        \textbf{Year} & \textbf{Credit card fraud reports} \\
        \hline
        2019 & 271,708 \\
        2020 & 393,440 \\
        2021 & 389,777 \\
        2022 & 440,666 \\
        2023 (Q1--Q3) & 318,087 \\
        \hline
    \end{tabular}
    \label{tab:fraud_by_year}
\end{table}

We present Table~2, showcasing credit card fraud reports categorized by the
United States. Credit card fraud dominated identity theft in 2022, with 440,666
reports and 318,087 reports filed in the first three quarters of 2023. Synthetic
fraud, the fastest-growing form of identity theft, led to \$1.8 billion in losses
for the auto lending industry in H1 2023, targeting retail and video game
industries. Individuals aged 30 to 39 reported the highest cases of identity
theft.

Georgia, Louisiana, and Florida ranked as the top three states for identity theft
per capita in 2022. Government documents or benefits fraud declined by 85\% in
2022 from 2021, with 57,912 reports filed compared to 396,025 in 2021. In 2022,
there were 1,802 data breaches, a 3\% decline from the all-time high in 2021,
impacting over 422 million people, up 44\% from 2021. To combat this, merchants
are advised to integrate advanced technologies like machine learning, AI,
risk-scoring, and behavioral analysis into their systems for comprehensive fraud
prevention [1].

\begin{table}[h]
    \centering
    \caption{\textbf{Credit card fraud in the United States}}
    \setlength{\tabcolsep}{6pt}
    \renewcommand{\arraystretch}{1.5}
    \begin{tabular}{|c|c|c|}
        \hline
        \textbf{Type of Identity} & \textbf{Reported cases, 2022} &
        \textbf{Percent change 2021--2022} \\
        \hline
        Credit Card Fraud       & 440,666 & 13\% \\
        Other Identity Theft    & 326,505 & 13\% \\
        Bank Fraud              & 156,134 & 25\% \\
        Loan or Lease Fraud     & 153,578 & 22\% \\
        Employment or Tax Fraud & 103,416 & 7\%  \\
        Phone or Utilities Fraud & 77,316 & 13\% \\
        Gov Doc or Benefits Fraud & 57,912 & 85\% \\
        \hline
    \end{tabular}
    \label{tab:fraud_us}
\end{table}

In 2022 and throughout the initial nine months of 2023, credit card fraud
maintained its status as the most prevalent form of identity theft. This trend
has been consistent, with credit card fraud ranking as the primary type of
identity theft from 2017 to 2019. However, there was a shift in 2020 and 2021
when government documents and benefits fraud surpassed credit card fraud, fueled
by opportunistic scammers taking advantage of pandemic-related government benefit
programs. Despite this temporary change, credit card fraud has demonstrated
persistent growth, experiencing only a minor 1\% decline in 2021. This decrease
followed a substantial 45\% surge from 2019 to 2020 and a noteworthy 72\%
increase from 2018 to 2019. The trend continued with a 13\% rise in 2022, and
preliminary reports from the first three quarters of 2023 suggest that credit
card fraud is maintaining a similar pace to that of 2022.

\begin{figure}[htp]
    \centering
    \includegraphics[width=8.5cm]{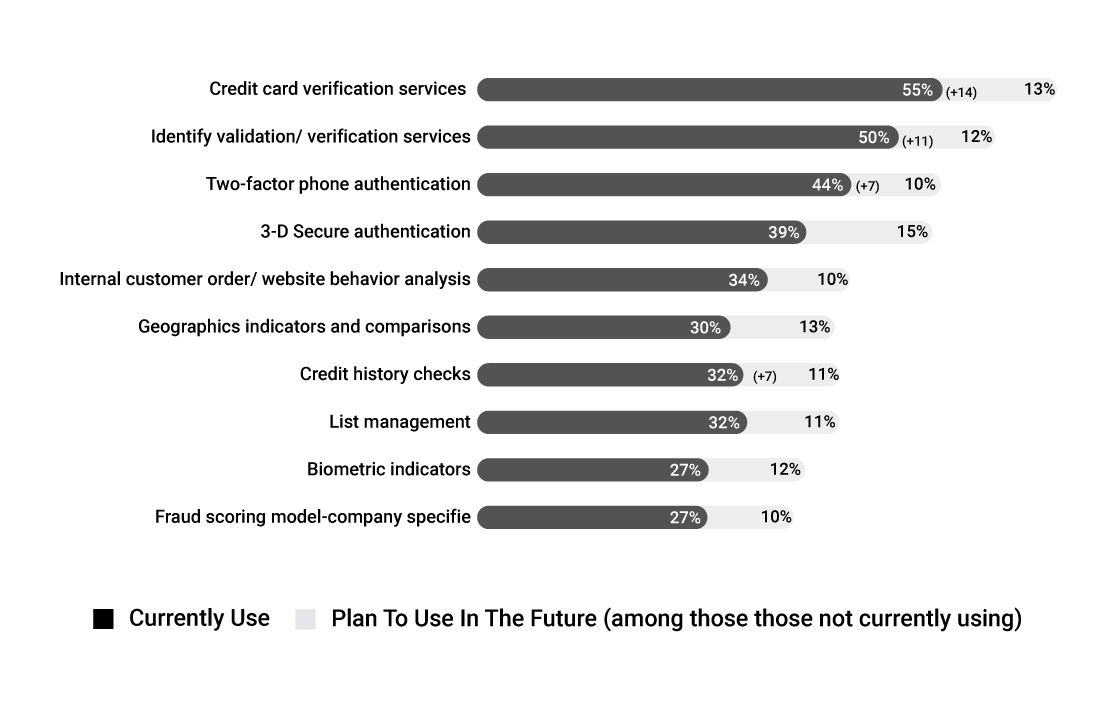}
    \caption{Current \& Future Usage of Fraud Detection Tools 2023}
    \label{fig:fig11}
\end{figure}

Merchants are increasingly incorporating a diverse range of tools for fraud
prevention, transitioning from strategic priorities to the tactical tools
essential for detecting and mitigating fraudulent activities. The survey indicates
a noteworthy rise in the average number of tools currently in use, increasing
from four in the previous year to five in the current year. Among the 15
different tools surveyed, eight have experienced a substantial increase in usage
by merchants on a global scale over the past year, as illustrated in
Figure~11 [20].

These tools encompass the three most widely adopted ones credit card verification
services, identity verification services, and two-factor phone authentication.
Additionally, a cluster of tools, utilized by approximately one-quarter to
one-third of all merchants, includes credit history checks, biometric indicators,
fraud scoring models, multi-merchant purchase velocity models, and order velocity
monitoring. Notably, the tools exhibiting the most significant surge in usage
over the past year align with those identified by a majority of non-users as
tools they intend to adopt in the future. This shift underscores a strategic
emphasis on enhancing fraud prevention measures through the incorporation of
proven and emerging tools within the merchant community.

\begin{figure}[htp]
    \centering
    \includegraphics[width=9cm]{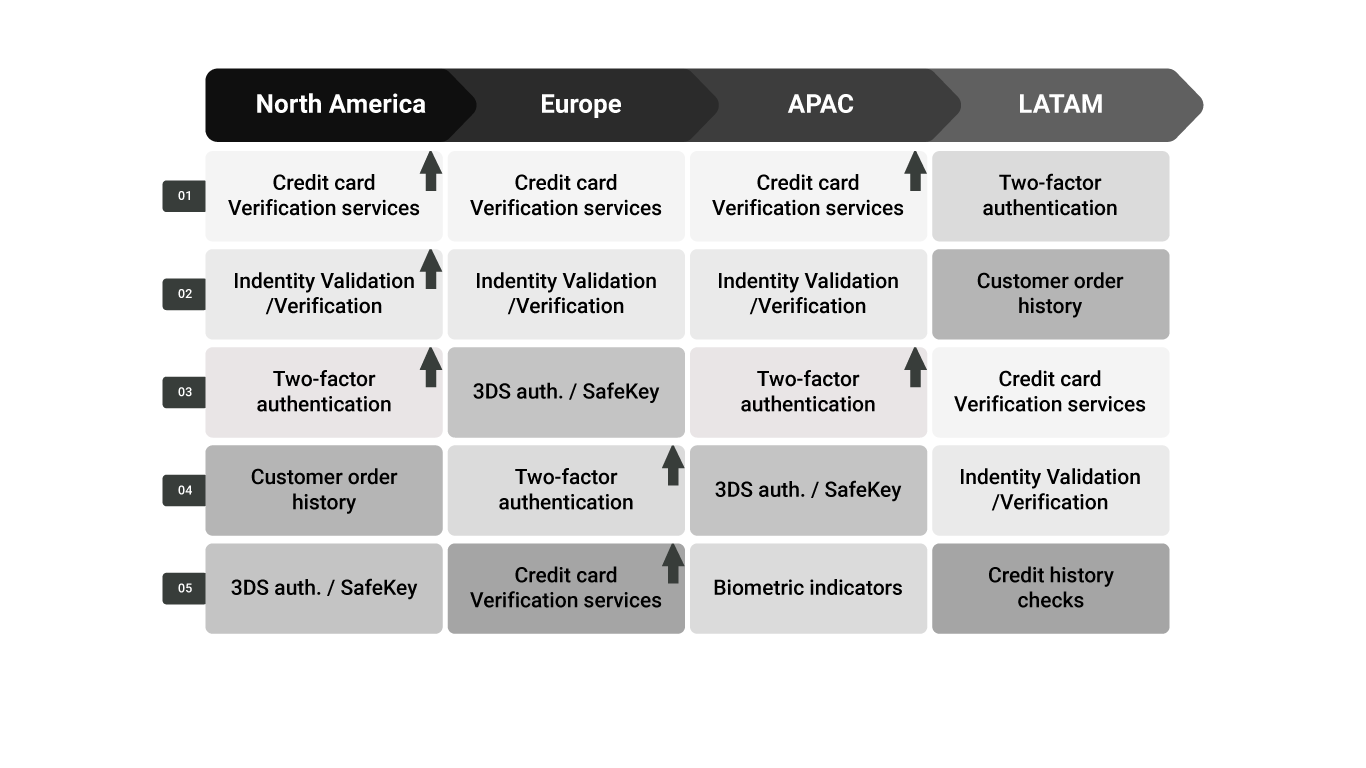}
    \caption{Top 5 Fraud Detection Tools Used By Region \& Size (2023)}
    \label{fig:fig12}
\end{figure}

In Figure~12, the top five tools employed by merchants in each region and size
category are delineated, along with the tools that witnessed a notably larger
adoption rate among merchants in each segment over the preceding year [20]. The
presence of green arrows in each column within the graphic signifies a substantial
increase in tool usage by merchants across nearly every segment. Particularly
noteworthy is the heightened adoption reported by merchants in North America and
enterprises, encompassing the majority of these top tools. The figures at the
bottom of Figure~11 draw attention to the discernible trend where merchants based
in APAC and LATAM, along with enterprises, exhibit a more extensive utilization
of tools compared to their counterparts in other segments.

This dataset in Figure~12 once again underscores a recent realignment in the
strategic approach of merchants, emphasizing the application of ``best practice''
tools such as two-factor authentication and identity verification. This shift
signals a move away from more intricate tools reliant on artificial intelligence
or machine learning algorithms, indicating a preference for established and
widely recognized tools for bolstering security measures.

\begin{figure*}[htbp]
    \centering
    \centerline{\includegraphics[width=1\textwidth]{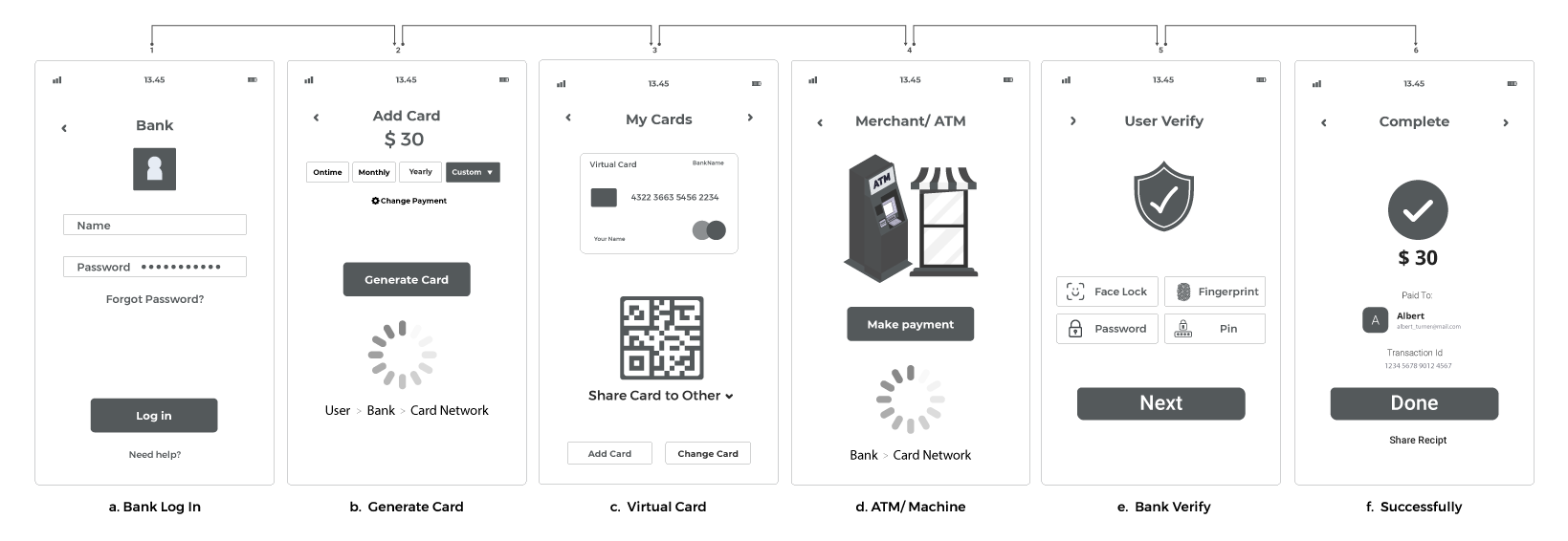}}
    \caption{Prototype for future cardless banking application}
    \label{fig:fig13}
\end{figure*}

In this research paper, we delve into the realm of modern banking security by
introducing a user prototype for a mobile banking application. Figure~13:
Cardless Banking App Screens. a)~Bank Login Screen: This interface serves as the
gateway for users to access the mobile banking app securely. Users are prompted
to input their designated username and password to log in, ensuring a
personalized and protected banking experience. b)~Generate Card Screen: Designed
to empower users with control over their virtual cards, this screen provides the
functionality to create a new virtual card. Users can customize the card by
setting spending limits and choosing the preferred card network. The intuitive
layout also incorporates a prominent ``Generate Card'' button for seamless card
creation. c)~My Cards Screen: Offering a comprehensive overview of the user's
virtual card portfolio, this screen displays vital card details such as the card
number, expiration date, and CVV code. Users can conveniently manage their
virtual cards with options to make payments, alter payment methods, share cards
with others, and add new cards to their collection.

In ``Merchant/ATM Selection'' (Figure~13d), users are presented with the choice
between ``Merchant'' and ``ATM'' options, allowing them to seamlessly navigate
between making purchases or withdrawing cash. This screen streamlines the
decision-making process for users based on their immediate financial needs.
Moving forward to ``User Verification'' (Figure~13e), this crucial screen prompts
users to verify their identity before completing a transaction. Utilizing
cutting-edge security measures, users can choose from Fingerprint, Face Lock, or
PIN options, ensuring a multi-layered authentication process for enhanced
security. Finally, in ``Confirmation and Actions'' (Figure~13f), users receive a
clear and concise confirmation message post-transaction. If the payment is
successful, the system confirms the order for merchant transactions or dispenses
cash for ATM withdrawals. This intuitive design aims to provide users with
real-time feedback on their actions, fostering a sense of confidence and
transparency in their mobile banking interactions. Through these additions to our
user prototype, we strive to create a secure, user-centric, and streamlined
mobile banking experience.

\section*{Discussion}
Securing bank card payments stands as a formidable barrier against the persistent
threat of sensitive financial data compromise. The ever-present risk of
interception or theft by malicious entities, such as hackers and fraudsters,
necessitates the implementation of robust security measures. In this context, a
secure bank card payment system emerges as a vital defense mechanism, offering
protection against identity theft and financial fraud. Beyond its role in
fortifying security, this approach introduces an additional layer of convenience
for consumers, liberating them from the burden of carrying physical cash or
managing multiple credit/debit cards for transactions. The proposed architecture
doesn't merely acknowledge the importance of secure bank card payments; it
pioneers a comprehensive methodology designed for the secure verification of
multiple elements within a novel bank payment system. This innovative
architectural framework meticulously delineates and verifies various components,
aiming to establish a resilient system that addresses multifaceted security
challenges. The focus on preventing fraud underscores a steadfast commitment to
creating a trustworthy and dependable framework for bank card transactions. By
encompassing these principles, the proposed architecture aspires to redefine the
standards of security and convenience in the realm of financial transactions.

\begin{table*}[t]
    \caption{\textbf{Comparison between the proposed architecture and other research}}
    \label{tab:comparison}
    \setlength{\tabcolsep}{6pt}
    \renewcommand{\arraystretch}{1.5}
    \begin{tabular}{m{0.17\textwidth} m{0.35\textwidth} m{0.40\textwidth}}
        \hline
        \textbf{Aspect} & \textbf{Proposed Architecture} & \textbf{Other Research} \\
        \hline
        New Architecture &
          Introduces a paradigm shift in approach to securing bank card transactions. &
          May lack emphasis on redesigning architecture for enhanced security. \\
        Card Number Generator &
          Enhances unpredictability of card numbers, mitigating risk of fraud. &
          Some approaches may overlook the importance of randomization in generating
          card numbers. \\
        Encrypted Card &
          Ensures confidentiality of sensitive information through advanced encryption
          techniques. &
          Encryption methods in other research may not offer the same level of
          protection for card details. \\
        Fraud Detection &
          Utilizes machine learning algorithms for proactive identification of
          fraudulent activities. &
          Other studies may rely solely on rule-based techniques, potentially missing
          emerging fraud patterns. \\
        User Verification &
          Enhances security through multi-factor authentication, fostering user trust. &
          User authentication methods in other research may not offer the same level
          of robustness against unauthorized access. \\
        \hline
    \end{tabular}
\end{table*}

1)~New Architecture: The introduction of the new architecture signifies a
paradigm shift in the approach to securing bank card transactions. By
meticulously outlining the system's components and verification processes, the
architecture establishes a robust foundation for enhanced security. This
innovative framework not only addresses current security challenges but is
designed with scalability to adapt to evolving threats, setting a new standard
in the realm of financial transaction security.

2)~Card Number Generator: The card number generator represents a critical
advancement in ensuring the uniqueness and security of generated card numbers.
By implementing sophisticated algorithms, the generator enhances the
unpredictability of card numbers, mitigating the risk of fraudulent activities.
This feature not only contributes to the prevention of unauthorized transactions
but also adds an extra layer of complexity, making it significantly challenging
for malicious actors to exploit vulnerabilities.

3)~Encrypted Card: The integration of encrypted cards adds a crucial layer of
protection to sensitive financial information. Employing advanced encryption
techniques ensures that card details remain indecipherable to unauthorized
entities, both during storage and transmission. This heightened level of
confidentiality safeguards against data breaches and unauthorized access,
enhancing the overall security posture of the banking card system.

4)~Fraud Detection: The incorporation of a sophisticated fraud detection system
is a pivotal aspect of the new architecture. Leveraging machine learning
algorithms and real-time transaction monitoring, the system can identify unusual
patterns and detect potential fraudulent activities. This proactive approach
enhances the system's ability to thwart fraudulent transactions, providing a
dynamic and adaptive defense against emerging threats in the ever-evolving
landscape of financial fraud.

5)~User Verification: The user verification component introduces an additional
layer of security by ensuring the legitimate identity of users. Through
multi-factor authentication and continuous verification mechanisms, the system
validates the authenticity of users, reducing the likelihood of unauthorized
access. This user-centric approach not only enhances security but also fosters
user trust and confidence in the banking card system, promoting a seamless and
secure user experience.

Table~3 provides a comparison between the proposed architecture and other
research in key aspects of securing bank card transactions. The proposed
architecture stands out for its innovative approach and comprehensive security
measures, including the introduction of a new architecture, a sophisticated card
number generator, encrypted card details, proactive fraud detection, and robust
user verification. In contrast, other research may focus on conventional methods
or overlook certain aspects critical for ensuring transaction security.

\section*{Conclusion and Future Work}
The surge in digital bank card payments necessitates a robust security framework.
This research proposes an advanced payment architecture, encompassing New
Architecture, Card Number Generator, Encrypted Card, Fraud Detection, and User
Verification. These components collectively fortify security in digital
transactions. The New Architecture redefines the foundation for secure digital
payments, while the Card Number Generator and Encrypted Card add layers of
sophistication to thwart fraud. The Fraud Detection system, driven by machine
learning, acts as a vigilant guardian. User Verification ensures identity
legitimacy through multi-factor authentication. Future efforts should prioritize
refining authentication, exploring encryption techniques, and advancing fraud
detection algorithms. As the digital payment market burgeons, ongoing
enhancements to the architecture and user awareness campaigns become paramount.
The proposed framework stands as a proactive approach, promising secure and
seamless digital payments in the evolving landscape. In summary, each component
within the discussed architecture contributes uniquely to fortifying the security
of bank card transactions. The synergistic integration of these elements reflects
a comprehensive and forward-thinking approach, aiming to create a banking card
system that not only meets current security standards but also anticipates and
addresses future challenges in the dynamic landscape of financial transactions.

\section*{Acknowledgment}
I would like to express my heartfelt appreciation to Dr.\ Varadraj Gurupur from
the Department of Electrical and Computer Engineering at the University of Central
Florida, Orlando, Florida, USA. His expertise and guidance have been
indispensable throughout the course of this research endeavor. Dr.\ Gurupur's
insightful feedback, constructive criticism, and unwavering support have played a
pivotal role in shaping the direction of our study and refining its methodology.
Furthermore, I am deeply grateful for Dr.\ Gurupur's willingness to share his
extensive knowledge and experience in the field. His mentorship has been
invaluable in navigating the complexities of our research topic and overcoming
various challenges encountered along the way. His commitment to academic
excellence and dedication to fostering a collaborative research environment have
been truly commendable. I would also like to acknowledge the Department of
Electrical and Computer Engineering at the University of Central Florida for
providing a conducive research environment and resources essential for the
successful completion of this study. The department's commitment to advancing
scientific knowledge and fostering innovation has been instrumental in
facilitating our research efforts. Overall, I am deeply thankful to Dr.\ Varadraj
Gurupur and all those who have contributed to this research in various capacities.
Their support, guidance, and encouragement have been instrumental in the
successful completion of this study.


\begin{IEEEbiography}[{\includegraphics[width=1in,height=1.25in,clip,keepaspectratio]{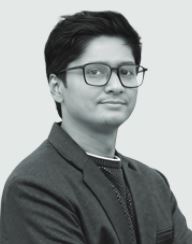}}]{Md. Israfeel}
Md Israfeel achieved his master's degrees in electrical and computer engineering
from the University of Central Florida, Florida, in 2024, and another master's
degree in computer science and engineering from Jahangirnagar University, Dhaka,
in 2022. Prior to this, he obtained his bachelor's degree in Computer Science and
Engineering from Daffodil International University, Dhaka, in 2020, and a Diploma
in Engineering in Computer Science and Technology from Bangladesh Sweden
Polytechnic, Rangamati, in 2012.

From 2016 to 2022, Md Israfeel was deeply immersed in Research and Development at
Alpha Net, Bangladesh, where he honed his skills and contributed to cutting-edge
projects. Since 2012, he has held the esteemed position of chief public lecturer
alongside the CEO at CSE Conference, Bangladesh. Notably, he is the Founder of
CSE Conference and ICT Bangladesh, showcasing his entrepreneurial spirit and
dedication to advancing the tech industry.

Md Israfeel's research interests span a wide range of topics, including Software
Design Development, IT Education Development, Artificial Intelligence, and Social
awareness based on Technology. His commitment to innovation and excellence has
been recognized through various accolades, including the 2020 Outstanding
Innovation and Motivation award from Sineris Web Services, FL, United States, and
the 2017 Best Entrepreneur and Innovation award from Daffodil International
University.
\end{IEEEbiography}


\begin{thebibliography}{00}
\bibitem{b1} Canadian Centre for Cyber Security, National Cyber Threat Assessment 2023--2024.
\url{https://www.cyber.gc.ca/sites/default/files/ncta-2023-24-web.pdf}

\bibitem{b2} Department of the Treasury, National Money Laundering Risk Assessment, February 2022.
\url{https://home.treasury.gov/system/files/136/2022-National-Money-Laundering-Risk-Assessment.pdf}

\bibitem{b3} Mastercard, Ecommerce fraud trends and statistics merchants need to know in 2024.
\url{https://b2b.mastercard.com/news-and-insights/blog/ecommerce-fraud-trends-and-statistics-merchants-need-to-know-in-2024/}

\bibitem{b4} merchantriskcouncil.org, 2023 Global Payments and Fraud Report.
\url{https://merchantriskcouncil.org/learning/mrc-exclusive-reports/global-payments-and-fraud-report}

\bibitem{b5} FFIEC Authentication and Access to Financial Institution Services and Systems Guidance, August 2021.
\url{https://www.ffiec.gov/press/pdf/Authentication-and-Access-to-Financial-Institution-Services-and-Systems.pdf}

\bibitem{b6} H.~Alamleh, A.~A.~S.~AlQahtani and B.~Al Smadi, ``Secure Mobile Payment Architecture Enabling Multi-factor Authentication,'' 2023 Systems and Information Engineering Design Symposium (SIEDS), Charlottesville, VA, USA, 2023, pp.~19--24, doi:~10.1109/SIEDS58326.2023.10137778.

\bibitem{b7} P.~Roy, P.~Rao, J.~Gajre, K.~Katake, A.~Jagtap and Y.~Gajmal, ``Comprehensive Analysis for Fraud Detection of Credit Card through Machine Learning,'' 2021 International Conference on Emerging Smart Computing and Informatics (ESCI), Pune, India, 2021, pp.~765--769, doi:~10.1109/ESCI50559.2021.9397029.

\bibitem{b8} A.~Koraus, J.~Dobrovi\v{c}, R.~Rajnoha, and I.~Brezina, ``The safety risks related to bank cards and cyber attacks,'' \textit{Journal of Security and Sustainability Issues}, vol.~6, pp.~563--574, 2017, doi:~10.9770/jssi.2017.6.4(3).

\bibitem{b9} M.~Al Rousan and B.~Intrigila, ``Multi-Factor Authentication for e-Government Services using a Smartphone Application and Biometric Identity Verification,'' \textit{Journal of Computer Science}, vol.~16, pp.~217--224, 2020, doi:~10.3844/jcssp.2020.217.224.

\bibitem{b10} J.~M.~Ashfield, ``Method and apparatus for using at least a portion of a one-time password as a dynamic card verification value,'' U.S. Patent 9,251,637, Feb.~2, 2016.

\bibitem{b11} G.~Ali, M.~A.~Dida, and A.~Elikana Sam, ``A Secure and Efficient Multi-Factor Authentication Algorithm for Mobile Money Applications,'' \textit{Future Internet}, vol.~13, no.~12, p.~299, 2021, doi:~10.3390/fi13120299.

\bibitem{b12} J.~S.~Kiernan and A.~Comoreanu, ``Credit Card Fraud Statistics,'' WalletHub, Oct.~2023.
\url{https://wallethub.com/edu/cc/credit-card-fraud-statistics/25725}

\bibitem{b13} J.~Caporal, ``Identity Theft and Credit Card Fraud Statistics for 2023,'' The Motley Fool.
\url{https://www.fool.com/the-ascent/research/identity-theft-credit-card-fraud-statistics/}

\bibitem{b14} B.~A.~Smadi, A.~A.~S.~AlQahtani and H.~Alamleh, ``Secure and Fraud Proof Online Payment System for Credit Cards,'' 2021 IEEE 12th Annual Ubiquitous Computing, Electronics \& Mobile Communication Conference (UEMCON), New York, NY, USA, 2021, pp.~0264--0268, doi:~10.1109/UEMCON53757.2021.9666549.

\bibitem{b15} C.~Gentry, ``Computing arbitrary functions of encrypted data,'' \textit{Communications of the ACM}, vol.~53, pp.~97--105, 2010.

\bibitem{b16} F.~Modler and M.~Kreh, \textit{Tutorium Analysis 1 und Lineare Algebra 1}. Springer: Berlin/Heidelberg, Germany, 2011.

\bibitem{b17} M.~Ogburn, C.~Turner, and P.~Dahal, ``Homomorphic Encryption,'' \textit{Procedia Computer Science}, vol.~20, pp.~502--509, 2013.

\bibitem{b18} M.~Schulze, \textit{Homomorphe Verschl\"{u}sselung und Europas Cloud}. Stiftung Wissenschaft und Politik, Berlin, Germany, 2021.

\bibitem{b19} F.~Armknecht et al., ``A Guide to Fully Homomorphic Encryption,'' Cryptology ePrint Archive, 2015.
\url{https://eprint.iacr.org/2015/1192.pdf}

\bibitem{b20} Cybersource, \textit{2023 Global Ecommerce Payments and Fraud Report}, 2023.
\url{https://www.cybersource.com/content/dam/documents/campaign/fraud-report/global-fraud-report-2023-en.pdf}

\bibitem{b21} Federal Trade Commission, ``New FTC Data Show Consumers Reported Losing Nearly \$8.8 Billion to Scams in 2022,'' Feb.~2023.
\url{https://www.ftc.gov/news-events/news/press-releases/2023/02/new-ftc-data-show-consumers-reported-losing-nearly-88-billion-scams-2022}

\bibitem{b22} Nilson Report, December 2021.
\url{https://nilsonreport.com/articles/card-fraud-losses-worldwide/}

\bibitem{b23} Statista, ``E-commerce payment fraud losses worldwide 2020--2023.''
\url{https://www.statista.com/statistics/1273177/ecommerce-payment-fraud-losses-globally/}

\bibitem{b24} Old National Bank, ``E-commerce fraud to cost \$48 billion globally as attacks skyrocket, report says.''
\url{https://www.oldnational.com/resources/insights/e-commerce-fraud-to-cost-48-billion-globally-this-year-as-attacks-skyrocket-report-says/}

\bibitem{b25} Juniper Research, ``Online Payment Fraud Losses to Exceed \$343 Billion Globally Over the Next 5 Years,'' 2022.
\url{https://www.juniperresearch.com/press/online-payment-fraud-losses-to-exceed-343bn/}

\bibitem{b26} Federal Deposit Insurance Corporation, \textit{Risk Review 2023}.
\url{https://fdic.gov/analysis/risk-review/2023-risk-review/2023-risk-review-full.pdf}

\bibitem{b27} E.~A.~L.~Marazqah Btoush et al., ``A systematic review of literature on credit card cyber fraud detection using machine and deep learning,'' \textit{PeerJ Computer Science}, vol.~9, p.~e1278, Apr.~2023, doi:~10.7717/peerj-cs.1278.

\bibitem{b28} Cybersource, \textit{Global Fraud and Payments Survey Report 2022}.
\url{https://www.cybersource.com/content/dam/documents/campaign/fraud-report/global-fraud-report-2022.pdf}

\bibitem{b29} S.~Mittal, P.~Jindal, and K.~R.~Ramkumar, ``Data Privacy and System Security for Banking on Clouds using Homomorphic Encryption,'' 2021 2nd International Conference for Emerging Technology (INCET), Belagavi, India, 2021, pp.~1--6, doi:~10.1109/INCET51464.2021.9456345.

\bibitem{b30} A.~Acar, H.~Aksu, A.~S.~Uluagac, and M.~Conti, ``A survey on homomorphic encryption schemes: Theory and implementation,'' \textit{ACM Computing Surveys}, vol.~51, pp.~1--35, 2018.

\bibitem{b31} Mass.gov, \textit{Handbook: Cybersecurity for the Financial Services Industry}, 2017.
\url{https://www.mass.gov/info-details/understand-cybersecurity-for-financial-institutions}

\bibitem{b32} U.S. Department of Justice, \textit{Financial Fraud in the United States}, 2017.
\url{https://bjs.ojp.gov/content/pub/pdf/ffus17.pdf}

\bibitem{b33} Juniper Research, \textit{Online Payment Fraud: Emerging Threats, Segment Analysis, and Market Forecasts 2022--2027}.
\url{https://www.juniperresearch.com/research/fintech-payments/fraud-identity/online-payment-fraud-research-report/}

\bibitem{b34} K.~W.~Hussein et al., ``Enhance Luhn algorithm for validation of credit card numbers,'' 2013.

\bibitem{b35} R.~Kiesel et al., ``Potential of Homomorphic Encryption for Cloud Computing Use Cases in Manufacturing,'' \textit{Journal of Cybersecurity and Privacy}, vol.~3, pp.~44--60, 2023, doi:~10.3390/jcp3010004.
\end{thebibliography}
\end{document}